\documentclass[prd,twocolumn,showpacs]{revtex4}
\usepackage{epsfig,amsmath,amssymb,mathtools,amscd}
\usepackage[active]{srcltx}
\usepackage[linkcolor=red]{hyperref} 
\usepackage{dsfont}
\usepackage{subfigure}
\newtheorem{lemma}{Lemma}

 \newtheorem{definition}{Definition}

\def\set#1{\mathrm #1}

\mathchardef\minus="002D 

\def\Proof{\medskip\par\noindent{\bf Proof. }}
\def\qed{$\,\blacksquare$\par}
\def\<{\langle}
\def\>{\rangle}
 \def\ket#1{| #1 \rangle} 
 
\def\ketbra#1#2{| #1 \rangle \langle#2 |}

\def\bvec#1{\boldsymbol{\mathrm #1}} 
\def\v#1{\boldsymbol{\mathrm #1}}

\def\Brill{\mathsf B}

\def\L2{{\mathcal L}_2}

\def\bk{\bvec{k}}

\newcommand{\dd}{\operatorname{d}}

\begin{document}
\title{Quantum walks, deformed relativity, and Hopf algebra symmetries} 
\author{Alessandro \surname{Bisio}} \email[]{alessandro.bisio@unipv.it} \affiliation{QUIT group, Dipartimento di
  Fisica, Universit\`{a} degli Studi di Pavia, via Bassi 6, 27100 Pavia, Italy}
\affiliation{Istituto Nazionale di Fisica Nucleare, Gruppo IV, via Bassi 6, 27100 Pavia, Italy}

\author{Giacomo Mauro \surname{D'Ariano}} \email[]{dariano@unipv.it}
\affiliation{QUIT group, Dipartimento di Fisica, Universit\`{a} degli
   Studi di Pavia, via Bassi 6, 27100
  Pavia, Italy}
\affiliation{Istituto Nazionale di
  Fisica  Nucleare, Gruppo IV, via Bassi 6, 27100
  Pavia, Italy}

\author{Paolo \surname{Perinotti}}
\email[]{paolo.perinotti@unipv.it} 
\affiliation{QUIT group, Dipartimento di Fisica, Universit\`{a} degli
   Studi di Pavia, via Bassi 6, 27100
  Pavia, Italy}
\affiliation{Istituto Nazionale di
  Fisica  Nucleare, Gruppo IV, via Bassi 6, 27100
  Pavia, Italy}
\begin{abstract}
We show how the Weyl quantum walk derived from principles in Ref.   \cite{PhysRevA.90.062106},
enjoying a nonlinear Lorentz symmetry of dynamics, allows one to introduce Hopf algebras for position
and momentum of the emerging particle. We focus on two special models of Hopf
algebras--the usual Poincar\'e  and the $\kappa$-Poincar\'e algebras. 
\end{abstract}
\maketitle

\section{Introduction}
%%%% Insert A head here
Quantum
walks (QWs)\cite{grossing1988quantum,meyer1996quantum,nayak2000quantum,ambainis2001one,aharonov2001quantum}
and more generally quantum cellular
automata (QCA)\cite{schumacher2004reversible,gross2012index,arrighi2011unitarity}
have been recently considered not only as a tool for quantum
simulation of
fields\cite{PhysRevA.73.054302,PhysRevA.81.062340,1751-8121-47-46-465302},
but also for the foundation of quantum field theory
\cite{bialynicki1994weyl,Yepez:2006p4406,darianopla,Bisio2015244,bisio2013dirac,PhysRevA.90.062106,bisio2014quantum,Bisio2015,Bisio2015-2}.
The  QCA framework appears as the natural candidate for the extension
of the informational paradigm, which has been crucial in the
understanding of foundations of Quantum Theory 
\cite{hardy2001quantum,dakic2009quantum,khrennikov2009contextual,PhysRevA.81.062348,masanes2011derivation,chiribella2011informational,Plotnitsky2015}),
to the foundation of Quantum Field Theory. 

The free theory has been derived starting from a denumerable set of
elementary quantum systems in interaction along with the general
assumptions of homogeneity, locality, isotropy, and linearity of the
interactions \cite{PhysRevA.90.062106,bisio2014quantum}. The whole
framework does not require Lorentz covariance, which results as a
subgroup of the dynamical symmetries of the quantum walk/automaton in
the limit of small
wave-vectors\cite{bibeau2013doubly,bisio2015lorentz}. For general
wave-vectors the Lorentz transformations are nonlinear, thus realizing
a model of Doubly Special Relativity
(DSR)\cite{amelino2001planck,PhysRevLett.88.190403,sym2010230}. 

In this paper we consider the simplest case of the mentioned quantum
walk field theory derived from principles, namely the one-particle
sector of the free Weyl automaton of
Ref. \cite{PhysRevA.90.062106}. We show how the dynamics of this walk
enjoys a nonlinear Lorentz symmetry, which allows us to introduce Hopf
algebras\cite{Lukierski1991331,majid2000foundations,kowalski2002doubly}
for position and momentum of the quantum walk particle,
generalizing the role of the Lie algebra of symmetries. We focus on two
special models of Hopf algebras: the usual Poincar\'e and the
$\kappa$-Poincar\'e algebras\cite{majid1994bicrossproduct}.

After reviewing the derivation of the Weyl quantum walk in
Sect. \ref{sec:weyl-automaton-3d} along with its symmetries, in
Section \ref{sec:from-poincare-hopf-1} we analyze the nonlinear
relativity symmetry, within the context of Hopf algebras---the
canonical framework in which deformed relativity models are studied
\cite{doi:10.1142/S0218271803003050,lukierski2003four,sym2010230}.  We
expound an analysis, closely related to the one in Ref.
\cite{kowalski2003non}, where we study how our non linear deformation
of the Lorentz group affects the Hopf algebraic construction of
spacetime and phase space.  We consider the nonlinear deformation in
the two alternative scenarios: the usual Poincar\'{e} and the
$\kappa$-Poincar\'{e} cases.  We will see that the construction of
spacetime as the dual space to the algebra of translations is left
unaffected by any nonlinear deformation that recovers the linear
Lorentz transformations at the leading order.  Whether we obtain the
usual spacetime or a noncommutative version is a feature that is
independent on the nonlinear transformation that we apply to the
momentum operators.  This is a slight generalization of the result of
Ref.\cite{kowalski2003non} where only the nonlinear deformations that
leave the rotation sector undeformed were considered.  On the other
hand, we see that the construction of the phase space as the left
cross-product algebra between momentum space and spacetime, does
depend on the nonlinear deformation. We then derive the set of
deformed Heisenberg commutation relations emerging in our framework
both in the usual Poincar\'{e} and in the $\kappa$-Poincar\'{e} cases.
Deformed Heisenberg commutation relations are an ubiquitous feature of
quantum Gravity models, they were first observed in the context of
String theory \cite{amati1987superstring,Gross1987129}, then studied
on their own right by many authors
\cite{PhysRevLett.101.221301,Scardigli199939,PhysRevD.52.1108,Maggiore199365,doi:10.1142/S0217751X95000085},
and recently considered for experimental verification
\cite{pikovski2011probing}.

\section{Quantum walk and relativity}
\label{sec:weyl-automaton-3d}

A quantum walk describes the discrete time evolution of particle on a discrete set $\Gamma$.  The
Hilbert space of the system is $\mathcal{H} := \ell^2(\Gamma) \otimes \mathbb{C}^s$ where
$\ell^2(\Gamma)$ is the Hilbert space of square summable function over $\Gamma$ and $\mathbb{C}^s$
is the Hilbert space corresponding to the internal degrees of freedom of the evolving particle.  We
introduce the ortonormal basis $\{\ket{g} \}$ of $\ell^2(\Gamma)$. The physical interpretation is
straightforward: the the state $\ket{g}\otimes\ket{\psi}$ correspond to a particle which is
localized in $g$ with internal state $\ket{\psi}$. The dynamics is described by a unitary operator
$A$ ($A^\dag A= A A^\dag =I$) on $\mathcal{H}$. As shown in Ref. \cite{PhysRevA.90.062106}, the
requirements of \emph{homogeneity} and \emph{locality} of the dynamics imply that the set $\Gamma$
is endowed with a graph structure orresponding to the Cayley graph of a group $G$\footnote{For the
  reader's convenience we remind the definition of Cayley graph. Let $G$ be a group and $S$ be a
  generating set of $G$.  The Cayley graph \unexpanded{$\Gamma=\Gamma(G,S)$} is a colored directed
  graph such that: $i)$ each element of $G$ corresponds to a vertex, $ii)$ each generator $S$ is
  assigned a color $c_s$,$iii)$ For any $g\in G$, $s\in S$, $g$ and $gs$ are joined by a directed
  edge of colour $c_s$.}. The generators of $G$ are represented by a translation operator $T_h$
acting on $\ell^2(\Gamma)$ as follows: $T_h \ket{g} = \ket{gh^{-1}}$ ($T$ is the right regular
representation of $G$).  Then, the homogeneity and locality assumption imply that the unitary
operator corresponding to the quantum walk $A$ can be decomposed as follows:
\begin{align}
  A = \sum_{h \in \set{S}} T_h \otimes A_h
\end{align}
where $\set{S} $ is the set of generatos and $A_h$ are operators on
$\mathbb{C}^s$.

Given a Cayley graph $\Gamma$ and a fixed dimension $s$ for the Hilbert space of the internal
degrees of freedom, the existence (or not) of a quantum walk on it is a highly nontrivial problem.
In Ref. \cite{PhysRevA.90.062106} some authors of the present manuscript addressed the case in which
$\Gamma$ is the Cayley graph of the Abelian group $\mathbb{Z}^3$ and the dimension of the internal
degree of freedom is $s=2$. Moreover, they assumed the quantum walk to be \emph{isotropic}, a
condition that translates the idea that all the directions on the lattice are equivalent. In
mathematical terms, there must exist a unitary representation $U$ over $\mathbb{C}^2$ of a group $L$
of graph automorphisms, transitive over a set of direct generators\footnote{The homogeneity
  assumption guarantees that the set \unexpanded{$\set{S}$} of generators can be split into disjoint
  subsets \unexpanded{$\set{S}_+ \cup e \cup \set{S}_-$} where \unexpanded{$\set{S}_-$} is the set
  of inverses of \unexpanded{$\set{S}_+$} and $e$ is the identity element.}, such that one has
$\sum_{h \in \set{S}} T_h \otimes A_h = \sum_{l(h) \in \set{S}} T_{l(h)} \otimes U_lA_hU^\dag_l$ for
all $l\in L$.  Under these assumptions, there is only one admissible Cayley graph of $\mathbb{Z}^3$,
which is the one corresponding to the body-centered cubic lattice, and there are only two admissible
quantum walks over it (up to a local change of basis).  The analytic expression of these quantum
walks are easily given in the Fourier transform basis $\ket{\bk} = (2 \pi)^{-3/2}\sum_{\bvec{x} \in
  \mathbb{Z}^3}e^{i \bk \cdot \bvec{x}} \ket{\bvec{x}}$ (where $\bvec{x}$ clearly denotes an element
in $\mathbb{Z}^3$)
\begin{align}
  \begin{split}\label{eq:weyl3D} 
    A^{\pm}& := \int_\Brill \!\!\dd \!\bk \ketbra{\bk}{\bk} \otimes
    A^{\pm}_{\bk}\\
    A^{\pm}_{\bk} &:= (2 \pi)^{-\frac32} \sum_{\bvec{y} \in \set{S}}
       e^{i\bk \cdot \bvec{y}} A^{\pm}_{\bvec{y}}\\
    A^{\pm}_{\bk} &:= \lambda^{\pm}(\bk)
    I-i{\bvec{n}}^{\pm}(\bk)\cdot\boldsymbol{\sigma}^\pm 
    \end{split}
\\
  \begin{split}
&{\bvec{n}}^{\pm}(\bk) :=
\begin{pmatrix}
s_x c_y c_z \pm c_x s_y s_z\\
c_x s_y c_z \mp s_x c_y s_z\\
c_x c_y s_z \pm s_x s_y c_z
\end{pmatrix},
\\
&\lambda^{\pm}(\bk) := (c_x c_y c_z \mp s_x s_y s_z ),\\
&c_\alpha := \cos({k}_\alpha/\sqrt{3}),\;s_\alpha:=
\sin({k}_\alpha/\sqrt{3}),\;\alpha = x,y,z.
\nonumber
  \end{split}
\end{align}
where $\Brill$ denotes the Brillouin zone of the body centered cubic lattice and
$\v\sigma^+=\v\sigma$ denote a vector of the usual Pauli matrices, while $\v\sigma^-=\v\sigma^T$
denotes the transposed ones.  The unitary constraint implies that $A^{\pm}_{\bk}$ is unitary for
every $\bk \in \Brill$. Notice that due to the discreteness of the lattice the quantum walk is
band-limited in $\bk$.  The quantum walk dynamics is determined by the solutions of the eigenvalue
equation $(A^{\pm} - e^{i \omega})\ket{\psi}=0$ that is equivalent to
\begin{align}
\label{eq:hamiltonian2}
(\sin\omega I -  \v{n}^{\pm}(\v{k})\cdot\v{\sigma}^{\pm})
\psi(\v{k} , \omega) =0,
\end{align}
which also implies the identity
\begin{align}
  \label{eq:disprel}
  \sin^2\omega-| \v{n}^{\pm}(\v{k})|^2 = 0
\end{align}
which defines the dispersion relation of the automaton.  It is easy to check that, by taking in the
limit $\bk \to \bk_0 = (0,0,0)$ in Eq. \eqref{eq:hamiltonian2}, the quantum walk $A^+$ (resp $A^-$)
recovers the dynamics of the right-handed (resp left-handed) Weyl equation.  Clearly, taking the
same limit in Eq. \eqref{eq:disprel} gives the usual relativistic dispersion relation $\omega^2 -
|\v{k}|^2 = 0$.  We notice that the same behaviour occurs in the limit $\bk \to \bk_2 =
\tfrac{\sqrt3\pi}{2}(-1,-1,-1)$ and in the limits $\bk \to \bk_1 = \tfrac{\sqrt3\pi}{2}(1,1,1)$,
$\bk\to \bk_3 = \sqrt3\pi(1,0,0)$ with the chirality exchanged.  Because of this reason we refer to
the quantum walks in Eq.~\eqref{eq:weyl3D} as \emph{Weyl walks}.  It is a remarkable result that a
Lorentz invariant dynamics is recovered from a dynamical model which follows from the only
assumptions of homogeneity, locality and isotropy, without the relativity principle.

In the following we will consider only the $A^+$ Weyl walk and we will drop the $\pm$ apex in
order to simplify the notation. The entire analysis can be straightforwardly applied to the $A^{-}$
case.

In the quantum walk framework space and time are not on an equal
footing:  space is given by the lattice structure, while time comes from
the discrete steps of the evolution. It is then far from obviuos
whether and how it is possible to recover changes of spacetime
coordinates that mix space and time, like boosts in special
relativity.  This question was recently addressed and answerd
in Ref. \cite{bisio2015lorentz} where the 
notion of change of observer for quantum
walks was defined as
as an invertible map  $\mathcal{L}_\beta$ over $[-\pi,\pi] \times
\Brill $, as follows
\begin{align}\label{eq:wk} 
(\omega, \bk) \to (\omega', \bk') =  \mathcal{L}_\beta (\omega, \bk) 
\end{align}
where the parameter $\beta$ labels different changes of reference-frame.  The idea is not to focus
on the discrete lattice coordinates and the discrete time step, but rather to consider $(\omega,
\bk)$--which are constants of motion of the quantum walk--as the fundamental variables.  In this
setting a symmetry of the dynamics is defined as follows:
\begin{definition}\label{def:symdynwalk}
 Let $A$ be a quantum walk on $\mathbb{Z}^3$.  A \emph{symmetry of the dynamics} for $A$ is a triple
$(\mathcal{L}_\beta, \Gamma_\beta, \tilde{\Gamma}_\beta)$, with $\mathcal{L}_\beta$ defined in
Eq~\eqref{eq:wk} and $\Gamma_\beta$, $\tilde{\Gamma}_\beta$ invertible matrix functions of $(\omega
,\bvec{k})$, such that
\begin{align}
\label{eq:invariantdynam}
(\sin\omega I -  \v{n}(\v{k})\cdot\v{\sigma})=
\tilde\Gamma^{-1}_\beta(\sin\omega' I -  \v{n}(\v{k'})\cdot\v{\sigma})\Gamma_\beta.
\end{align}
The set of symmetries $\mathbf{S}^A$ is a group which we refer to as the \emph{symmetry group of the
quantum walk} $A$. 
\end{definition}
The next step is then to explore whether the symmetry group of the Weyl walk $A$ contains a
representation of the Lorentz group which recovers the usual one in the regime in which the walk
approaches the Weyl equation (i.e. near $\bk_0 ,\bk_1,\bk_2 $, and $\bk_3$ ).  In other words we are
asking whether there exists a \emph{deformed relativity model} which preserves the dynamics of the
Weyl walk $A$.

Deformed (or doubly) special relativity is a theoretical proposal in which one modifies the linear
Lorentz transformations in order to have an invariant energy scale in addition to the speed of
light. Such a theory has been proposed by Amelino-Camelia\cite{amelino2001planck} and developed by
other authors\cite{PhysRevLett.88.190403} as a kinematic structure which may underlie quantum theory
of gravity.  Indeed, if the Planck length were a threshold beyond which quantum gravity effects
would become relevant, this length should be the same for all the observers, a statement which
clearly disagrees with special relativity.  A deformed relativity model consist in replacing the
usual (linear) Lorentz transformation $L_\beta$ in momentum space as follows:
\begin{align}
  \begin{split}\label{eq:defrelgen}
     L_\beta \to \mathcal{L}_\beta, \\
\mathcal{L}_\beta = \mathcal{D}^{-1} \circ L_\beta \circ \mathcal{D},\\
(\omega, \bk ) \to  \mathcal{L}_\beta (\omega, \bk ), 
  \end{split}
\end{align}
where the map $\mathcal{D}$ is a singular invertible map such that its Jacobian $J_{\mathcal{D}}$
equals the identity in $(\omega,\v{k})=0$.  These conditions are needed in order to have an
invariant energy, while recovering the usual phenomenology at energy scales much smaller than the
Planck scale.

For a complete derivation where we refer to Ref \cite{bisio2015lorentz}. Apart from a null measure
set we split the Brilloun zone $\Brill$ into four parts $\Brill_i$, $i=0, \dots 3$. Each vector
$\bk_i$ belongs to the corresponding region $\Brill_i$. The regions $\Brill_i$ are chosen such that
the compositions $\mathcal{L}^{(i)}_\beta = \mathcal{D}^{(i)-1} \circ L_\beta \circ
\mathcal{D}^{(i)}$ are well defined, with $\mathcal{D}^{(i)}$ given by

\begin{align}
\label{eq:theultimatedeformation2}
\begin{aligned}
 & \mathcal{D}^{(i)} :  \mathsf{\Sigma}_i \to \mathsf{\Gamma}_0, \qquad
  \mathcal{D}^{(i)} : 
\begin{pmatrix}
 \omega\\
\v{k}
\end{pmatrix}
 \mapsto 
g(\omega, \bk)
\begin{pmatrix}
  \sin\omega\\
\v{n}^{(i)}(\v{k})
\end{pmatrix},\\
&\mathsf{\Sigma}_i := \{ (\omega , \v{k}) \text{ s.t. }   \v{k} \in
\Brill_i , \sin^2\omega - |\v{k}|^2 = 0 \},\\
& \mathsf{\Gamma}_0 := \{ p\in \mathbb{R}^4 \mbox{ s.t. }  p_\mu p^\mu =0\},
  \end{aligned} 
  \end{align}
  for a suitably defined function $g(\omega, \bk)$ \footnote{An admissible expression of the
    function \unexpanded{$g(\omega, \v{m})$} is explicitly given in Ref.  \cite{bisio2015lorentz}.
    For the following consideration it suffices to know that \unexpanded{$g(\bvec{\bk_i}) = 1$} and
    \unexpanded{$\nabla g(\bvec{\bk_i}) = \bvec{0}$ for all $i=0, \dots, 3$}.}.  The maps
  $\mathcal{L}^{(i)}_\beta$ provide a well defined nonlinear representation of the Lorentz group on
  each set $\Sigma_i$.

  For $i=0,2$ one can easily check that the conditions of Definition~\ref{def:symdynwalk} are met if
  we set $\Gamma_k = \Lambda_\beta$ and $\tilde{\Gamma}_k = \tilde{\Lambda}_\beta$, provided that
  $\Lambda_\beta$ is the right handed spinor representation of the Lorentz group, and
  $\tilde{\Lambda}_\beta$ is the left-handed representation.  For $i=1,3$ the same holds provided we
  exchange the two representations. The four vector $(\omega, \bk) \in \mathsf{\Sigma}_i$ transforms
  under the nonlinear representation $\mathcal{L}^{(i)}_\beta$.  Since $\cup_{i=0}^3\Brill_i=\Brill$
  (apart from a zero-measure set), we have that the maps $\mathcal{L}^{(i)}_\beta$ provide a notion
  of Lorentz transformation for any solution of the Weyl QCA dynamics.

  We notice that the choice of the map \eqref{eq:theultimatedeformation2} is not unique, since there
  are many admissible choices for the function $g(\omega, \bk)$.  The symmetry group $\mathbf{S}^A$
  of the Weyl walk $A$ contains then many different istances of deformed relativity. However, all of
  them will recover the usual Lorentz transformations near the points $\bk_i$. The four invariant
  regions are interpreted as four different particles (this is the phenomenon of Fermion doubling).

  Finally, it is worth stressing the reversed perspective of this approach with respect to the usual
  one in relativistic quantum mechanics. The Weyl walk dynamics has been singled out without
  requiring Lorentz invariance, whereas the Lorentz invariance is recovered as a symmetry of the
  dynamics.

\section{Hopf Algebra, $\kappa$-Poincare and noncommutative spacetime}

\label{sec:from-poincare-hopf-1}

In this section we explore how the deformation of the Lorentz group
given by the nonlinear deformation \eqref{eq:theultimatedeformation2}
manifests itself at the level of the Poincar\'{e} algebra. We will
restrict to the $\mathcal{D}^{(0)}$ case and then drop the $\,^{(0)}$
apex in order to simplify the notation, the generalization for
$i = 1,2,3$ is trivial. In order to perform this analysis we will need
to consider the framework of Hopf algebras (for a comprehensive
introduction to the subject we suggest
Ref.\cite{majid2000foundations}). The notion of Hopf algebra
generalizes that of Lie algebra to a less ``rigid'' object, which is
can accommodate a nonlinear version of the Lorentz group, which is
incompatible with a Lie algebra structure. Unfortunately, any specific
nonlinear deformation of the Lorentz group, of the kind in
Eq. \eqref{eq:defrelgen}, is not sufficient to select a unique Hopf
algebra, since there are many compatible coproduct
structures. Nevertheless it is interesting to study the role that our
deformed Lorentz transformation plays within the context of Hopf
algebras, since this is the canonical context in the specialized
literature on deformed relativity
\cite{doi:10.1142/S0218271803003050,lukierski2003four,sym2010230}.

\subsection{Classical Poincar\'{e} and $\kappa$-Poincar\'{e} Hopf algebras }

The Lie algebra of the Poincare group
is given by the relations 
\begin{align}
  \begin{aligned}
\label{eq:liepoincare}
    [M_i,M_j] &= i\epsilon_{ijk} M_k &   [M_i,p_j] &= i\epsilon_{ijk} p_k \\
  [M_i,N_j] &= i\epsilon_{ijk} N_k &[M_i,p_0] &= 0\\
  [N_i,N_j] &= -i\epsilon_{ijk} M_k &[N_i,p_j] &= i\delta_{ij} p_0 \\
  [N_i,p_0] &= -ip_0 &  [p_\mu,p_\nu] &=0    
  \end{aligned}
\end{align}
where we denoted with $M_i$ the generators of spatial rotations, with
$N_i$ the generators of boosts, and with $p_\mu$ the generators of
translations---$p_0$ denoting the generator of time
translation. Clearly, if we apply a non-linear map to the generators
$p_\mu$, the set of commutation relations \eqref{eq:liepoincare} is
spoiled, and generally does not define a Lie algebra anymore. However,
it is possible to treat such deformations on formal grounds, within
the more general setting of Hopf algebras. The universal enveloping
algebra of the Lie algebra \eqref{eq:liepoincare} can be endowed with
a Hopf algebra structure by defining the primitive co-product
$ \Delta $, antipode $S$, and co-unit $\epsilon$ as
\begin{align}
\label{eq:poincarehopf}
  \begin{aligned}
    \Delta ( O )&= 1 \otimes O + O \otimes 1,\\
  S (O) &= -O, 
  &S (1) &= 1,\\
  \epsilon (O) &=0,
  &\epsilon (1) &= 1.
  \end{aligned}
\end{align}
These relations are just a rephrasing of the usual Poincar\'{e} Lie
algebra structure \eqref{eq:liepoincare} in the language of Hopf
algebras, where the additional coalgebra structure allows one to
express the Leibniz rule for the infinitesimal action of the group on
products of functions through the coproduct. This rule can be easily
accounted for using the tensor product structure and the theory of
group representations. On the other hand, within the context of Hopf
algebras any invertible analytical map that transforms momenta as
$p'_\nu = f_\nu (p_\mu)$ can be treated as a change of basis in an
infinite dimensional algebra. Even if, from a mathematical
perspective, this transformation is just a change of basis, it may
have significant physical consequences like e.g.~a deformation of the
dispersion relation.

Nonlinear modifications of the translation generators is not the only
possible deformation of the classical Poincar\'{e} symmetry. It is
indeed possible to consider scenarios in which the Hopf-algebraic
structure itself is different (up to any change of basis) from the
classical one given by Eqs.~\eqref{eq:liepoincare} and
\eqref{eq:poincarehopf}. Of particularly interest are those
deformations of the classical Poincar\'{e} Hopf algebra that reduce to
the usual one in a suitable limit of values of the deformation
parameters. The classification of all the possible deformation of
Poincar\'{e} Hopf algebra is still an open problem.

Up to now the most studied example is the so-called
$\kappa$-Poincar\'{e} Hopf algebra
\cite{majid1994bicrossproduct,Lukierski1991331}, which in the so
called ``classical basis''
\cite{:/content/aip/journal/jmp/34/12/10.1063/1.530247,doi:10.1142/S0218271803003050}
takes the following form:
\begin{align}
\label{eq:kpoincare}
\begin{aligned}
  & \mbox{the same algebraic sector} \\
&\begin{aligned}
  \Delta (p_0) &= \frac{\kappa}{2} (K \otimes K - K^{-1} \otimes
 K^{-1}) +\\
&+\frac{1}{2 \kappa} (K^{-1} |p|^2 \otimes K^{-1}) \\
&+ (K^{-1} p_i \otimes p_i+K^{-1} \otimes K^{-1} |p|^2) \\
\end{aligned}\\
& \Delta (p_i) = p_i \otimes K + 1 \otimes p_i 
\end{aligned}
\end{align}
where $K := \frac{1}{\kappa} (p_0 + (p_0^2 - |p|^2 + \kappa^2)^{\frac12})$
and $\kappa$ is a real parameter. One can check that the usual classical  Poincar\'{e} Hopf algebra is recovered in the limit $\kappa \to \infty$.

Then, starting from the enveloping algebra of the Poincar\'{e} Lie
algebra we have two different roads that can be explored: i) assume
the coalgebra structure \eqref{eq:poincarehopf} and consider the
classical Poincar\'{e} Hopf algebra, or ii) assume
Eq. \eqref{eq:kpoincare} and study the $\kappa$- Poincar\'{e} Hopf
algebra.  On one hand, our scenario singles out a set of generators
$k_\mu$ that are defined in terms of the classical one $p_\mu$ by the
nonlinear deformation $p = \mathcal{D}(k)$.  On
the other hand, our model does not prefer any of the different
algebric models and it is interesting to consider the consequences of
the the nonlinear deformation given by the map $\mathcal{D}$ in both
the classical Poincar\'{e} and in the $\kappa$- Poincar\'{e} cases.

\subsection{From Poincar\'{e} Hopf algebra to spacetime}
\label{sec:from-poincare-hopf}
One of the most popular speculations concern the relation between the
algebra of position coordinate and the algebra of translation.

If we denote by $T$ the Hopf algebra generated by the translation
generators $p_\mu$
one can define the position algebra as the dual hopf algebra $T^*$
on which $T$ acts covariantly \cite{majid1994bicrossproduct}.
$T^*$ is determined by introducing the
generators $x_\mu$ and 
the pairing
\begin{align}
\label{eq:pairing}
   \<f(p_\mu), x_\nu \> = f(\frac{\partial}{\partial x_\mu})[x_\nu] (0).
\end{align}
This way of introducing the pairing follows the classical pairing
between the enveloping algebra of $\mathbb{R}^4$ with the algebra of
functions on $\mathbb{R}^4$, i.e.~the translation generators act as
derivatives evaluated at the origin. The structure of $T^*$ is then
determined by the axioms of Hopf algebra duality
\begin{align}
  \begin{aligned}
     \<p, xy \> = \<\Delta(p), x \otimes y \> \\
 \< pq, x \> = \<p \otimes q, \Delta(x) \>.
  \end{aligned}
\end{align}
Since the momenta commute we have that positions cocommute
with co-commutators
\begin{align}
 \Delta{x_\mu} =  1 \otimes x_\mu + x_\mu \otimes 1.
\end{align}

The commutation relations $[x_\mu, x_\nu]$ are different from $0$ only
if the coproducts for the $p_\mu$ are not co-commutative.  Then, if we
are dealing with the usual Poincar{\'e} algebra we will always have a
commutative spacetime, independently of the nonlinear mapping we are
using to define the generators, as their coproduct will still be
co-commutative.

The scenario is different in the $\kappa$-Poincar\'{e} case where it
has been proved that the Hopf algebra defined by
Eqs. \eqref{eq:kpoincare}
leads to the following commutation relations for positions
\begin{align}
  \label{eq:kminkovski}
  [x_i,x_j]=0 \quad [x_0,x_i] = -\frac{i}{\kappa} x_i
\end{align} In this case it could happen that a differrent choice of the generators $p_\mu$ could lead to different commutation relations.  In the literature \cite{kowalski2003non} it is proved that the commutation relations \eqref{eq:kminkovski} do not depend on the choice of basis as long as it is rotationally invariant and such that the usual generators are recovered in the limit $\kappa \to \infty$. It is possible to slightly generalize this result by dropping the assumption of rotational invariance
\begin{lemma}\label{lem:kappa}
  Let $\mathcal{M}:p \mapsto p' = \mathcal{M}(p)$ be a transformation
  of the translation generators such that $J_\mathcal{M}(0)= I$. Then
  the commutation relations \eqref{eq:kminkovski} remain unchanged.
\end{lemma}
\Proof
First we observe that, from the pairing \eqref{eq:pairing} we have
that the only terms in the cocommutators \eqref{eq:kpoincare}
that are relevant for computing  the commutators $[x_\mu, x_\nu]$ are the ones that are at most
bilinear, i.e. 
 $\Delta (p_0) = 1 \otimes p_0 + p_0 \otimes 1 + \frac{1}{\kappa}\sum_ip_i
 \otimes p_i$ and 
 $ \Delta (p_i) = p_i \otimes 1 + \frac{1}{\kappa} p_i
 \otimes p_0 + 1 \otimes p_i$. 
By power expanding $\mathcal{M}$ we have
$p'_\mu = p_\mu + \frac{1}{\kappa} m_{\alpha \beta} p_\alpha p_\beta$
and by  power expanding the inverse function $\mathcal{M}^{-1}$ we have
$p_\mu = p'_\mu + \frac{1}{\kappa} n_{\alpha \beta} p'_\alpha
p'_\beta$.
It is then easy to verify that, up to the bilinear terms, the coproduct $\Delta (p'_0)$ is co-commutative while the coproducts $\Delta (p'_i)$ are the sum of a co-commutative term and $\frac{1}{\kappa} p'_i\otimes p'_0$. Since the non-cocommutative term $\frac{1}{\kappa} p'_i\otimes p'_0$ has the same expression independently of the nonlinear mapping $\mathcal{M}$, the commutation relation for the spacetime variables remains the same.\qed
This result tells us that our nonlinear mapping, which  
satisfies the hypotheses of lemma \ref{lem:kappa}, does not change the commutation relations for the spacetime variables.

\subsection{From Poincar\'{e} Hopf algebra to phase space}
We have seen in the preceding section that a notion of spacetime can
be introduced as the dual $ T^* $  to the Hopf algebra of translations
$T$.
The additional notion of  
\emph{left coregular action} 
\begin{align}
  p \rhd x := \< p , x_{(2)} \> x_{(1)}
\end{align}
allows to introduce a notion of phase space\cite{doi:10.1142/S0218271803003050,amelino1997kappa} as 
the \emph{left cross product} algebra $T^* \rtimes T$
where the multiplication is defined as
\begin{align}
  (x \otimes p)(x' \otimes p') = x(p_{(1)} \rhd x') \otimes p_{(2)} p'.
\end{align} 
If we define the isomorphisms
\begin{align}
  x \sim x \otimes 1 \quad 
  p \sim 1 \otimes p \quad 
\end{align}
it make sense to consider the commutation relation
\begin{align}
  \begin{split}
\label{eq:commutphase}
     [p_\mu, x_\nu] &= x_\nu \otimes p_\mu - \< p_{\mu(1)} , x_\nu \> 1 \otimes p_{\mu(2)} - \\
& + \< p_{\mu(1)} , 1 \> x_\nu \otimes p_{\mu(2)} \\
  \end{split}
\end{align}
We will see that the commutation relations 
\eqref{eq:commutphase} will depend on the choice of the generators,
i.e. they depend on the nonlinear deformation.

We will now compute the commutation relation
\eqref{eq:commutphase} for the choice of generators
given by the map $\mathcal{D}$ . Since we cannot derive an analytic
expression for the inverse map
$\mathcal{D}^{-1}$ we will consider just the terms up to the first
order in $\frac1\kappa$.
We have then
\begin{align}
  \label{eq:dlcnbasis}
  \begin{aligned}
    E     & =   \omega  \\
p_x  & =   k_x        +   \frac{1}{\kappa} k_y k_z  \\
p_y   &=    k_y  -       \frac{1}{\kappa} k_x k_z   \\
p_z  & =   k_z         + \frac{1}{\kappa} k_x k_y 
  \end{aligned}
\qquad \quad
  \begin{aligned}
    \omega  &    =   E  \\
k_x  & =   p_x        -  \frac{1}{\kappa} p_y p_z \\
k_y  & =    p_y  +  \frac{1}{\kappa} p_x p_z \\
k_z  & =   p_z         -  \frac{1}{\kappa} p_x p_y  
  \end{aligned}
\end{align}
This result holds the same for any choice of $g(\omega,\bk)$
such that $\nabla g(\bvec{0})=\bvec{0} $.

After some cumbersome but straightforward calculation,
we have, in the classical
Poincar\'{e} Hopf algebra case
\begin{align}
  \label{eq:dlcnphaseclas}
  \begin{aligned}
&[k_i , x_j] = -i \delta_{ij} -i \frac{(-1)^{\delta_{i,2}}}{\kappa}
\left( \delta_{i+1,j}k_{i+2} + \delta_{i+2,j}k_{i+1} \right)   \\
&[\omega, x_j] = [k_i,t] = 0 \qquad
[\omega, t ] = i 
  \end{aligned}
\end{align}
where we used the notation $x=1, y=2,z =3$ and the sums are meant 
to be modulo $3$.
Similarly in the 
$\kappa$-Poincar\'{e} Hopf algebra case
we get
\begin{align}
  \label{eq:dlcnphasekappa}
  \begin{aligned}
&
\begin{aligned}
[k_i , x_j] =& -i \delta_{ij}(1-\frac{\omega}{\kappa})+ \\
& -i \frac{(-1)^{\delta_{i,2}}}{\kappa}
\left( \delta_{i+1,j}k_{i+2} + \delta_{i+2,j}k_{i+1} \right)
\end{aligned}
\\
&[\omega, x_j] = \frac{i}{\kappa} k_j
- \frac{1}{2 \kappa} x_j |k|^2 \qquad
 [k_i,t] = 0 \\
&[\omega, t ] = i   - \frac{1}{2 \kappa} x_j |k|^2
  \end{aligned}
\end{align}

Differently from the space-time commutation relations,
the commutation
relation between position and momentum are affected by
the choice of the basis. 
As one could expect, in both cases we recover the usual 
commutation relations between position and momentum
as the deformation parameter $\kappa$ goes to infinity.

\section{Conclusion}
In this paper we have studied the dynamical symmetries of the Weyl
quantum walk. As explained in the paper such walk is particularly
interesting since it was derived from general principles without
assuming Lorentz covariance, but nevertheless it recovers a
Lorentz-invariant dynamics in the limit of small wave-vectors. For
large wave-vectors the Lorentz group becomes nonlinear, and we have a
model of Doubly Special Relativity. We introduced the Hopf algebras
for position and momentum of the quantum walk particle, and evaluated
the structure constants of the algebras for the usual Poincar\'e and
the $\kappa$-Poincar\'e cases. Generalizing a result of
Ref.\cite{kowalski2003non}, we have shown that the spacetime
commutators are left unaffected by any nonlinear deformation that
recovers the linear Lorentz transformations at the leading
order. Finally we derived the analytical expression up to the first
order in the inverse Planck-energy $\kappa^{-1}$ of the deformed
Heisenberg commutation relations.

\acknowledgments This work has been supported by the Templeton Foundation under the project ID\#
43796 {\em A Quantum-Digital Universe}.  

\bibliographystyle{apsrev4-1}
\bibliography{bibliography}

\end{document}